\documentclass[sigconf, citation-style=numeric]{acmart}
\AtBeginDocument{%
  }
\usepackage{algorithm}
\usepackage{algorithmic}
\usepackage{caption}

\setcopyright{none}
\renewcommand\footnotetextcopyrightpermission[1]{}
\newcommand{\fptc}{\textsc{fptc}}

\usepackage{mdframed}
\usepackage{xcolor}
\usepackage{graphicx}
\newmdenv[
  linewidth=0.8pt,
  linecolor=black,
  backgroundcolor=gray!10,
  innerleftmargin=8pt,
  innerrightmargin=8pt,
  innertopmargin=6pt,
  innerbottommargin=6pt,
  skipabove=\baselineskip,
  skipbelow=\baselineskip,
]{observationbox}

\newcounter{observation}
\newcommand{\observation}[1]{%
  \begin{observationbox}
    \noindent\textbf{Observation \Roman{observation}}\stepcounter{observation}: #1
  \end{observationbox}
}
\begin{document}

\fancyhead[RO,LE]{}


\title[\fptc{}: A Fast Parallel Transform-based Codec for Efficient Asymmetric Signal Compression]{FPTC: A Fast Parallel Transform-based Codec \\\ for Efficient Asymmetric Signal Compression}

\author{Ben Mechels}
\affiliation{%
  \institution{University of Minnesota}
  \city{Minneapolis}
  \state{MN}
  \country{USA}
}
\email{meche046@umn.edu}

\author{Ryan Billmeyer}
\affiliation{%
  \institution{Medtronic}
  \city{Minneapolis}
  \state{MN}
  \country{USA}
}
\email{ryan.billmeyer@medtronic.com}

\author{Alexander Chen}
\affiliation{%
  \institution{University of Minnesota}
  \city{Minneapolis}
  \state{MN}
  \country{USA}
}
\email{chen8206@umn.edu}

\author{Shiyang Li}
\affiliation{%
  \institution{University of Minnesota}
  \city{Minneapolis}
  \state{MN}
  \country{USA}
}
\email{li004074@umn.edu}

\author{Caiwen Ding}
\authornote{Corresponding author.}
\affiliation{%
  \institution{University of Minnesota}
  \city{Minneapolis}
  \state{MN}
  \country{USA}
}
\email{dingc@umn.edu}

\renewcommand{\shortauthors}{Mechels et al.}

\begin{abstract}

Modern high-performance computing and Internet-of-Things deployments increasingly generate large volumes of signal data that must be compressed efficiently on resource-constrained acquisition devices and decompressed at scale on centralized servers. Lossy compression is widely adopted to minimize storage and transmission costs on low-power hardware sensors, yet existing methods rarely optimize for both reconstruction quality and decompression throughput simultaneously, nor do they apply methods that generalize across signal domains. In this work, we introduce \fptc{}, a high-throughput asymmetric signal codec that pairs a lightweight sequential encoder with a massively parallel GPU decoder designed for server-side batch decompression. \fptc{} applies a windowed discrete cosine transform (DCT) to exploit frequency-domain sparsity, quantizes spectral coefficients with a hybrid three-zone mapping, and entropy codes the result using Huffman coding with a novel packing scheme. The pipeline used in \fptc{} is designed to be throughput oriented on the GPU, maximizing performance without sacrificing reconstruction quality. We evaluate \fptc{} on ten datasets spanning four signal domains: biomedical diagnostic, seismic reflections, power-grid production metrics, and meteorological recordings. Our results demonstrate that \fptc{} outperforms existing frameworks in compression ratio while maintaining competitive throughput, achieving multiplicative compression performance of 3.6x (power), 3.1x (meterological), 1.5x (biomedical), and 1.2x (seismic) over existing frameworks.

\end{abstract}

\maketitle

\fancyhead[L]{\shorttitle}
\fancyhead[R]{\shortauthors}

\section{Introduction}
\label{sec:intro}

Today’s high-performance computing (HPC) and Internet-of-Things (IoT) deployments generate massive volumes of heterogeneous data from a wide range of sensing modalities. These streams are acquired continuously on resource-constrained devices, thus transmitting or storing raw data is often impractical under limited bandwidth, memory, and energy budgets \cite{ImplantSurvey2008,zhang2012compressed}. In these domains, data is compressed and sent to centralized servers for archival, reconstruction, filtering, analytics, or machine-learning pipelines. Compression is therefore a fundamental requirement for making such sensing systems practical end to end. However, while these workloads primarily take the form of sampled time-series, they are highly heterogeneous in amplitude range, temporal smoothness, correlation structure, noise characteristics, and tolerance to distortion \cite{AFC-2024,Srisooksai-WSN-Survey-2012}. This heterogeneity makes compression fundamentally challenging: a method tuned for one data family often generalizes poorly to others, either sacrificing compression ratio or inducing unacceptable reconstruction error.

\begin{figure}[H]
    \centering
    \includegraphics[width=1\columnwidth]{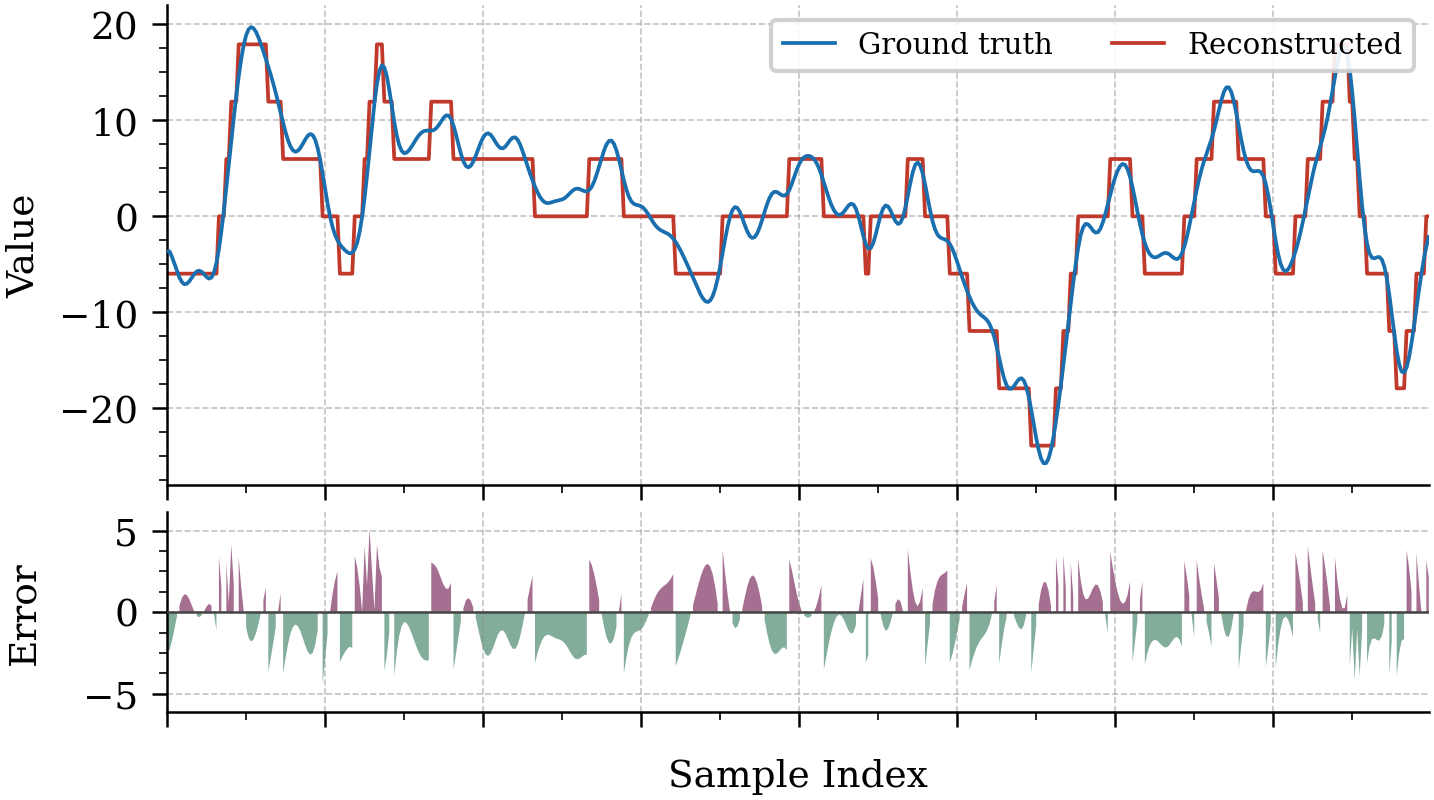}
    \caption{Prediction-based HPC Compressors~\cite{cuSZp22024} exhibit major distortion compressing EEG waveforms at 10x CR.}
    \label{fig:compression-error}
\end{figure}

Lossy compression is the predominant strategy for meeting these constraints. By tolerating bounded reconstruction error, lossy codecs achieve compression ratios beyond what is practical with lossless methods, reducing data transmission energy and storage cost. However, existing compressors often optimize only a subset of the design objectives. Recent work has highlighted a trade-off between compression and runtime: many compressors that achieve strong compression ratios do so at low throughput, whereas many high throughput codecs provide substantially weaker compression effectiveness~\cite{RECOIL2024}. On the decoding side, recent work has demonstrated that general-purpose compressed formats can be decoded at high throughput on massively parallel hardware such as GPUs~\cite{lin2023recoil}, but such designs are not tailored to the statistical structure and distortion requirements of transformed signal data. Another class of compressors, especially in biomedical settings, primarily optimize compression ratio (CR) and distortion with percentage root-mean-square difference (PRD), while giving much less attention to throughput-oriented parallel decompression~\cite{blanco2005prdcr,dasan2021ecg,zheng2021ecg}. Few existing systems jointly optimize for \emph{both} a minimal resource encoder and a high throughput decoder, and fewer still consider signal fidelity as a primary design objective alongside compression and throughput.

\begin{figure}[t]
      \centering
      \includegraphics[width=1\linewidth]{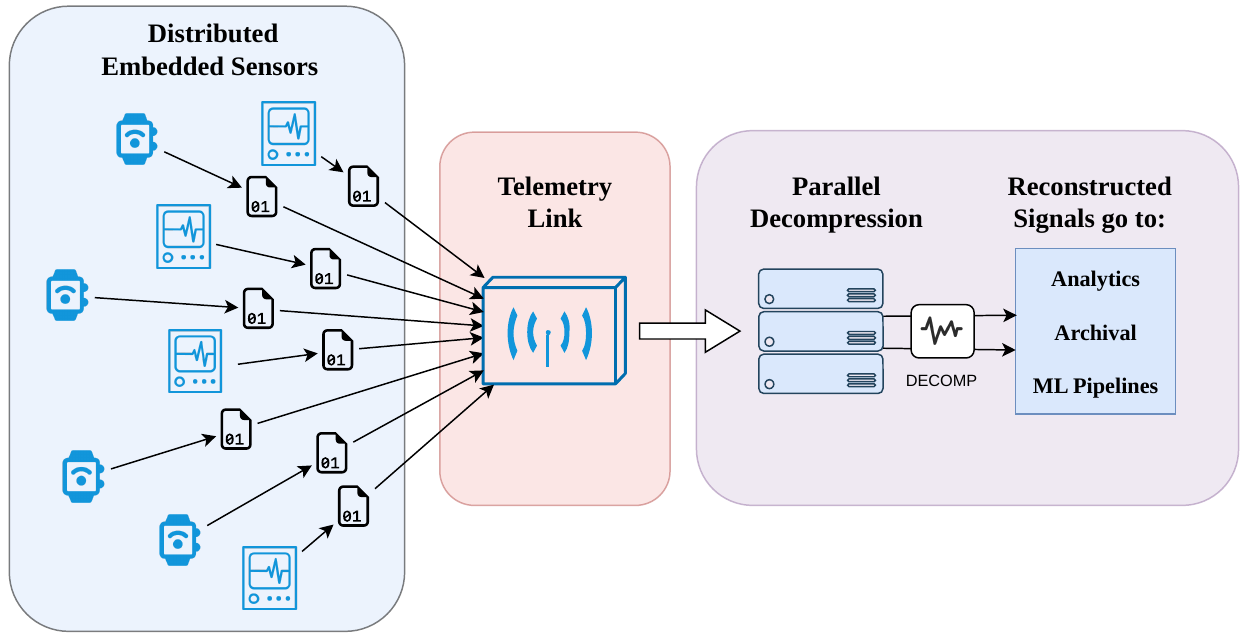}
      \caption{\fptc{} asymmetric system architecture}
      \label{fig:FPTC System Diagram}
\end{figure}

In this work, we introduce \fptc{}, an asymmetric lossy codec designed for signal domains where compression runs on resource-constrained embedded hardware and decompression runs at scale on GPUs. The pipeline of the codec is tuned through parameters and prebuilt structures. In a given domain, representative datasets are utilized to precompute data structures for the most algorithmically challenging parts of the pipeline: the quantization table and the Huffman coding tree. These structures are deployed per signal domain to provide low-complexity encoding at scale and high throughput decompression on centralized GPU servers. Our contributions are as follows:

\begin{itemize}

    \item We present \fptc{}, an asymmetric lossy signal codec designed for sensing workloads, with a lightweight sequential encoder for resource-constrained devices and a massively parallel GPU decoder for high-throughput server-side reconstruction.

    \item We introduce a transform-based compression pipeline that combines windowed DCT, hybrid three-zone quantization, and length-limited canonical Huffman coding, while using offline-trained quantization tables and codebooks to reduce runtime encoder complexity.

    \item We design a \emph{SymLen} bitstream format that packs Huffman codewords into self-contained words with per-word symbol counts, enabling low-overhead, fully independent GPU Huffman decoding without inter-thread synchronization.

    \item We develop a dual-fused GPU decompression pipeline. The first stage kernel fuses per-thread Huffman decoding with variable buffer compaction, and the second stage kernel fuses dequantization with inverse DCT reconstruction using a uniform per-sample work assignment.

\end{itemize}

 We evaluate \fptc{} across 4 signal domains, including biomedical, geophysical, infrastructure, and meteorological data, and show that it outperforms existing technologies in compression, throughput, and reconstruction fidelity.
 
\section{Related Work}

This section outlines existing design objectives in codec pipeline designs in both low-power embedded and massively parallel HPC systems and how they differ. Primarily, the objectives are divided into complexity-focused embedded compression and throughput-focused HPC big data workflows.

\subsection{Embedded Compression}

Many sensing deployments follow an asymmetric execution model: data is acquired continuously on resource-constrained devices, then transmitted to centralized servers for archival, reconstruction, and downstream analysis. This pattern appears in wireless sensor networks, wearable and implantable biomedical devices, smart-health systems, and other cloud telemetry settings~\cite{Sadler2006SLZW,Marcelloni2009TinyNodes,Duarte2012WSNCompression,Razzaque2012WSNSurvey,Abdellatif2019EdgeHealth}. In such systems, battery life, memory capacity, and compute budget are tightly constrained, making low-complexity on-device compression a critical consideration. The severity of these constraints are often understated in the literature, especially for continuous signal workloads. For example, ECG and EEG monitoring produce long data streams that are expensive to transmit or archive in raw form, particularly in wearable and implantable devices ~\cite{ImplantSurvey2008}. 

Furthermore, many studies on physiological signal compression focus on CR and PRD with limited consideration for energy and compute constraints. These methods commonly use complex transforms, adaptive quantization, dynamic thresholding, and elaborate entropy coding methods to optimize CR and PRD for a specific signal class~\cite{ECGDWTHuffman2021,ECGPWM2024,ECGPerf2015,ECGDynamicThreshold2024,ECGSelfControlled2024,ECGEfficient2023}. While effective, such methods can be impractical for resource-constrained embedded systems. Similar challenges arise in seismic acquisition, power-grid telemetry, and environmental sensing. Across these domains, efficient compression is not simply a storage optimization but a requirement for real deployment~\cite{Energy-aware-compression, Sadler2006SLZW}. Furthermore, these domains exhibit local correlation in the frequency-domain, making transform-based lossy compression attractive. Prior work in signal compression and wireless sensing has shown that transform-domain representations can reduce redundancy while preserving fidelity~\cite{ECGDWTHuffman2021,ECGEfficient2023}. This motivates a design that utilizes a common transform and entropy-coding framework with established efficient implementations, while allowing signal-class-specific tuning.

\subsection{Parallel Compression}

\begin{figure*}[!t]
    \centering
    \hspace{-8mm}
    \includegraphics[width=1\linewidth]{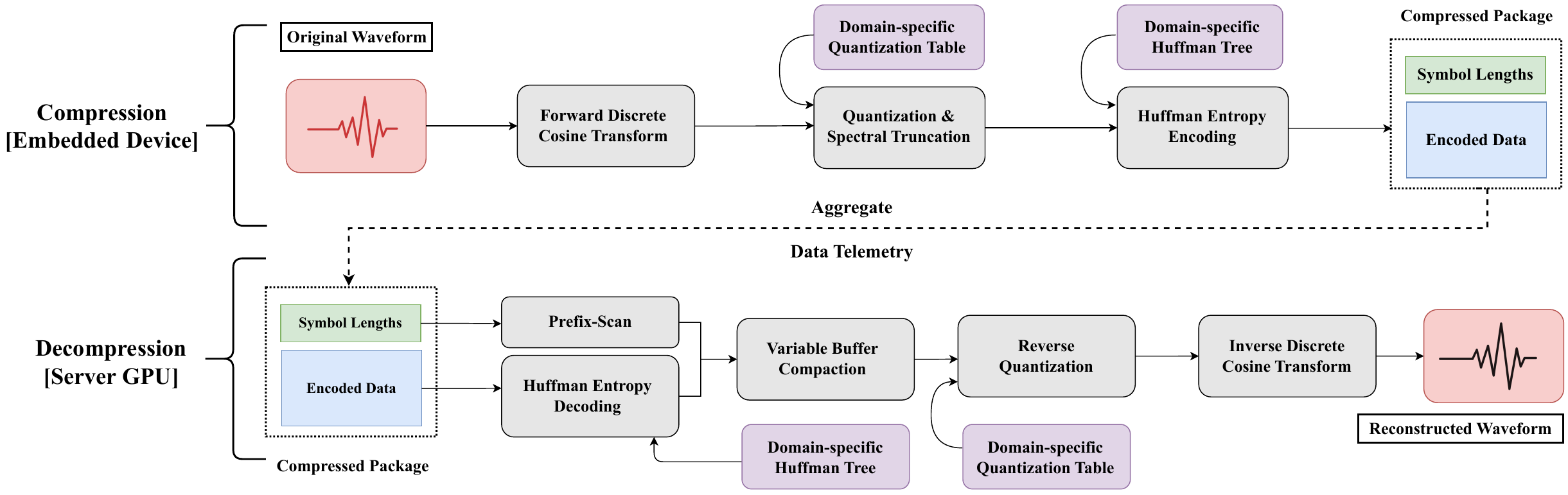}
    \caption{\fptc{} compression and decompression pipelines}
    \label{fig:pipe}
\end{figure*}

Recent work in high-performance computing has produced several GPU-oriented lossy compressors that target strong rate-distortion performance and high throughput on floating-point scientific data, including cuSZp~\cite{cuSZp2023}, cuSZp2~\cite{cuSZp22024}, FZ-GPU~\cite{FZGPU2023}, PFPL~\cite{PFPL2025}, and CUDA-enabled ZFP~\cite{ZFP}. These systems are most relevant for our baselines on centralized parallel decompression as they prioritize compression ratio with bounded error with foremost interest in extreme throughput on GPU accelerators. However, these methods target general-purpose floating-point compression using prediction-based models, and do not exploit the transform strategy used in \fptc{}. Parallel Huffman decoding has attracted significant attention due to its widespread use and lossless nature. Prior work has explored massively parallel decode strategies, self-synchronizing streams, gap-array methods, canonical-Huffman decoding on many-core processors, and GPU-oriented decompression ~\cite{MassivelyParallelHuffman2018,GapArrays2020,CanonicalManyCore2020,ParallelJPEGHuffman,RECOIL2024}. These works highlight the primary challenges in developing high-throughput Huffman decoders: the variable-length nature of codewords creating thread divergence and the indeterminability of codeword positions. Domain-specific GPU decompression work has explored these challenges in seismic processing using wavelet and Huffman based methods~\cite{SeismicWaveletGPU2015,SeismicHuffmanGPU2015}.

\section{Design Overview}

\fptc{} is structured around an asymmetric model motivated on each end by the constraints of signal acquisition systems and the enormous compute of modern GPUs. In domains such as medical devices, insertable cardiac monitors have very limited computation and energy budgets. This presents the need for on-board compression to be a low-complexity, single pass pipeline. After aggregate data from many cardiac monitors arrive on centralized servers, GPU architectures can be utilized to decompress this data simultaneously at high throughput. Figure~\ref{fig:pipe} illustrates the end-to-end data flow. The codec has three compression stages: transform, quantization, and entropy coding.

\subsection{Transform Stage}

Time to frequency transformation is the first step of \fptc{}. While this transform does compress some data itself under certain parameters, it mainly serves to prepare signal strips for compression in subsequent stages by exploiting frequency sparsity. For signals, images, audio, and many more applications this technique is well studied \cite{lawson2002jpeg2000}. In our pipeline, each signal strip of length $S$ is partitioned into non-overlapping windows of $N$ samples. Within each window, the signal is transformed using the type-II discrete cosine transform (DCT-II). For a window of samples $x[0], x[1], \ldots, x[N{-}1]$, the forward DCT computes $N$ frequency-domain coefficients:

\begin{equation}
\label{eq:dct}
C[k] = \frac{2}{N} \sum_{n=0}^{N-1} x[n] \cos\!\left(\frac{\pi}{N}\left(n + \tfrac{1}{2}\right) k\right), \quad k = 0, 1, \ldots, N{-}1.
\end{equation}

The DCT is chosen over the discrete Fourier transform (DFT) for two main reasons. First, the DCT produces real-valued coefficients, halving the storage requirement compared to complex DFT output. Second, the  even-symmetric extension of the DCT avoids boundary discontinuities that would spread energy across many bins \cite{khayam2003discrete}, resulting in better energy compaction for our signal applications.

The key property exploited by \fptc{} is \emph{frequency sparsity}: for the signal classes of interest, the vast majority of spectral energy concentrates in a small number of low-frequency coefficients. Given a window of size $N$, only the first $E \leq N$ coefficients are retained and the remaining $N - E$ high-frequency bins are discarded before quantization. This spectral truncation achieves an immediate $N/E$ reduction in the symbol count entering the entropy coder, while the energy compaction property shows that the discarded bins will contribute negligibly to reconstruction quality. The truncation boundary $E$ is a parameter that is set per signal class based on the smoothness and sampling rate of that class.

\subsection{Quantization Stage}

After spectral transform, each retained coefficient is a 32-bit float. The quantization stage maps these to 8-bit unsigned integers, achieving a set $4{\times}$ compression ratio. Rather than applying a single quantization rule uniformly, \fptc{} uses a \emph{hybrid three-zone} quantizer that adapts its mapping to the statistical properties of each frequency band within the DCT window. The $E$ retained coefficient indices from the DCT output are partitioned into three contiguous zones with two boundaries $B_1$ and $B_2$.

\subsubsection{Zone~0: $\mu$-Law Companding (bins $0$ to $B_1{-}1$)}

The lowest-frequency bins carry the most energy and exhibit the widest dynamic range, thus they must be preserved carefully. These coefficients are quantized with $\mu$-law companding, a non-linear mapping that allocates finer quantization resolution near zero and progressively coarser resolution toward the extremes \cite{Gray-Neuhoff-Quantization}. This is a direct tradeoff where the output of this stage will exhibit a more uniform distribution less suited for compression in entropy coding but we find that the reconstruction fidelity it provides makes this tradeoff worthwhile. For a coefficient $c$ with per-zone maximum $A_0$, the compressed value is:

\begin{equation}
\label{eq:mulaw}
q = \frac{\ln(1 + \mu \cdot |c|/A_0)}{\ln(1 + \mu)}, \quad |c| \leq A_0
\end{equation}

\noindent where $\mu$ is a configurable companding parameter set per signal class that controls the degree of non-linearity. The compressed value $q \in [0,1]$ is then mapped to an 8-bit level: positive coefficients occupy bins $129$--$255$, negative coefficients occupy bins $0$--$127$, and the zero bin is fixed at $128$. The maximum $A_0$ is computed as a clipped percentile of the absolute coefficient values across all windows at the given frequency bands, rejecting outliers that would otherwise waste quantization levels on rare extremes. Importantly, since the distributions of randomly sampled DCT windows are very similar, we are able to make assumptions about the DCT output that allow more fine tuned quantization compared to frameworks without transforms.

\subsubsection{Zone~1: Linear Deadzone (bins $B_1$ to $B_2{-}1$)}

Mid-band coefficients carry moderate energy. These are quantized with a symmetric linear mapping enhanced by a \emph{deadzone}, a region around zero within which all values are collapsed to the zero bin. The deadzone ratio $\alpha_1$ is a tunable constant. Increasing $\alpha_1$ forces more coefficients to the zero bin, improving entropy coding efficiency at the cost of reconstruction fidelity. For a coefficient $c$ with per-zone maximum $A_1$ and deadzone width $d_1 = \alpha_1 A_1$

\begin{equation}
\label{eq:deadzone}
q = \begin{cases}
129 + \left\lfloor \dfrac{c - d_1}{A_1 - d_1} \cdot 126 + 0.5 \right\rfloor & \text{if } c > d_1, \\[6pt]
127 - \left\lfloor \dfrac{|c| - d_1}{A_1 - d_1} \cdot 127 + 0.5 \right\rfloor & \text{if } c < -d_1, \\[6pt]
128 & \text{otherwise.}
\end{cases}
\end{equation}

\subsubsection{Zone~2: Aggressive Zeroing (bins $B_2$ to $E{-}1$)}

High-frequency coefficients carry minimal signal energy. The deadzone in Zone~2 spans the entire dynamic range and \emph{every coefficient maps to the zero bin}. This concentrates quantized values onto a single symbol, thereby reducing the average codeword length. Although Zone~2 coefficients carry no information post-quantization, they are still present in the bitstream when $B_2 \le E{-}1$ as highly compressible values.

\subsection{Entropy Coding Stage}

The quantized coefficient stream is entropy coded with a length-limited canonical Huffman scheme ~\cite{LL-Huffman}, compressing 1-byte symbols into variable-length codewords. This Huffman variant enables low-complexity, low-memory embedded encoding and high-throughput, independent decoding. A symbol histogram is first computed on the quantized coefficients of a representative dataset. From this histogram, optimal code lengths are determined based on a maximum length constraint $L_{\max}$ using the Larmore--Hirschberg package-merge algorithm which solves the length-limited minimum-redundancy coding problem in $O(\sigma L_{\max})$ time for an alphabet of size $\sigma$, which is 256 based on 1-byte values encoded post-quantization.

The length-limited variant of Huffman coding is employed to bound the runtime memory usage of the encoder and decoder. A lookup-table (LUT) with $2^{L_{max}}$ entries is used to achieve $O(1)$ conversions. Bounding the LUT size allows it remain cache-resident during both compression and decompression. Futhermore, this codebook is canonical as the Larmore–Hirschberg package merge algorithm provides a symbol to codeword length mapping, thus codes are canonized iteratively to provide codeword bit mappings.

\subsection{Signal Domain Parameters}

Signal-domain parameters are determined by evaluating codec performance on a representative dataset, optimizing for reconstruction quality while throughput remains a secondary consideration. Parameter choices significantly effect reconstruction fidelity but have limited impact on throughput. Table~\ref{tab:params} provides the \fptc{} parameters with typical values for reference along with the ranges tested. Note that not all values in this table vary signficantly, such as $\alpha_1$ for example, which remains relatively constant across all testing.

\begin{figure}[t]
      \centering
      \hspace{-4mm}
      \includegraphics[width=1.04\linewidth]{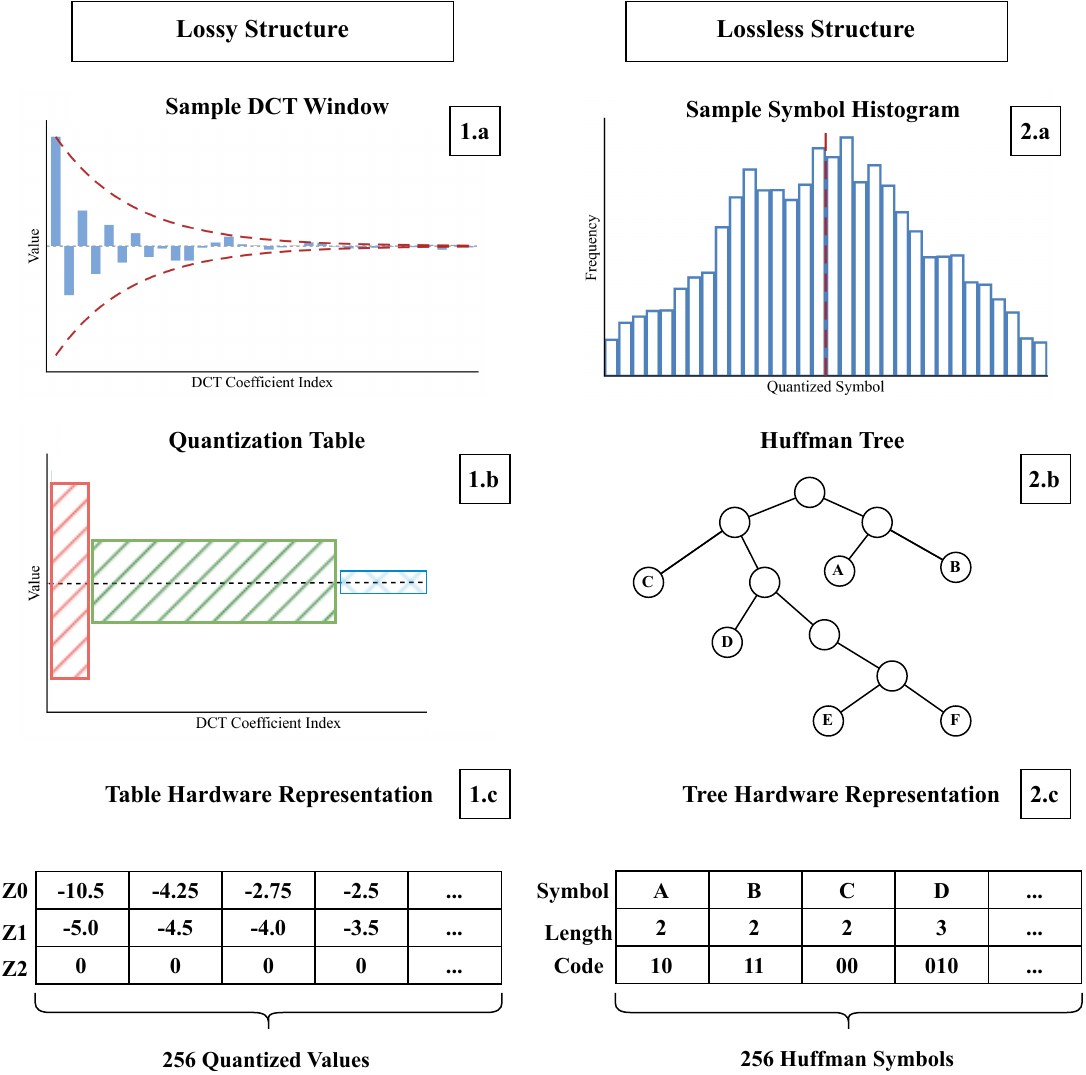}
      \caption{Codec structures, including quantization table (1.b) built from DCT data (1.a) and represented as a multidim array (1.c), and Huffman tree (2.b) built from a coefficient histogram (2.a) and represented as a flattened tree (2.c).}
      \label{fig:structure_diagram}
\end{figure}

\subsubsection{Lossy} We find that the typical values for $\mu$ and $\alpha_1$ are performant across all studied signal domains while parameters $N, E, B_1$ and $A_0$ exhibit higher sensitivity to the application signal domain. $N$ and $E$ relate to the local smoothness and sampling rate of a signal. For instance, we observe that Load Power is highly smooth and thus necessitates a lower $E:N$ ratio for optimal compression in comparison to the less structured EEG. $B_1$ is determined by rate of decay of the average coeffcient window. $A_0$ relates to the variance between the probability distributions of distinct DCT windows. For example, the sinus rhythm of ECG data affords the ability to only $\mu$-compand the very first coefficient, where EEG requires this zone to cover more coefficients for fidelty. For similar reasons, the percentile clip $A_0$ is set higher on ECG since the domain has high stationarity.

\subsubsection{Lossless} The entropy coding stage uses a pre-computed codebook for Huffman look-up operations. The parameters in Huffman are more nuanced as the tree-building itself is exact to a certain distribution. This codebook can be constructed from general \fptc{} DCT coefficient distributions of the specific domain, but when applied to randomly sampled domain data, it will only approximate the most optimal length-limited encoding of that data. However, this is an intrinsic property of Huffman regardless \cite{LightweightHuffman}. We find an approxmation in this way is more than sufficient to achieve high compression due to signal stationarity.

\begin{table*}[t]
\caption{Dataset parameters with typical values on studied signal classes.}
\label{tab:params}
\centering
\begin{tabular*}{\textwidth}{@{\extracolsep{\fill}} l c l c c @{}}
\toprule
\textbf{Parameter} & \textbf{Symbol} & \textbf{Description} & \textbf{Range} & \textbf{Typical Value} \\
\midrule
\texttt{DCT\_SIZE}           & $N$        & Transform block size (DCT dimension) & $[4, 128]$        & 32 \\
\texttt{ENCODED\_COEFFS}     & $E$        & Number of retained low-frequency coefficients & $[1, N]$        & 16 \\
\texttt{HYBRID\_BOUNDARY\_1} & $B_1$      & Boundary separating low- and mid-frequency regions & $[0, E]$         & 2 \\
\texttt{HYBRID\_BOUNDARY\_2} & $B_2$      & Boundary separating mid- and high-frequency regions & $[B_1, E]$        & 16 \\
\texttt{MU\_COMPANDING}      & $\mu$     & Quantization or scaling strength parameter & $[1, 500]$       & 50.0 \\
\texttt{DEAD\_RATIO\_ZONE1}  & $\alpha_1$ & Deadzone ratio for mid-frequency band & $[0.0, 1]$    & 0.004 \\
\texttt{ZONE\_PERCENTILE}   & $A_0$       & Percentile threshold for zone bounds & $[90.0, 100.0]$  & 99.9 \\
\bottomrule
\end{tabular*}
\end{table*}

\section{Optimization}

The following section covers how \fptc{} is implemented from the proposed design in Section 3. Where the design overview focuses on how the proposed design enables high compression with reconstruction fidelity, this section focuses on the hardware realization of the codec. We first describe the compression pipeline, emphasizing lowing algorithmic complexity, then present the decompression pipeline with a focus on exploiting parallelism on GPUs.

\subsection{Compression}

To enable single pass encoding, the encoding side of \fptc{} uses a prebuilt quantization table and Huffman codebook. The encoder executes three steps of transformation according to these structures. First, it computes the DCT as discussed in Section 3.1. It then converts spectral coefficients from their \texttt{float} values to quantized \texttt{uint8} types according to the quantization table. Finally, the encoder must deal with alignment in converting the predictable length input in fixed byte lengths to variable bit-length codes packed into words. We introduce a \emph{SymLen} method that cleanly packs the variable-sized compressed bitstream of Huffman encoding into fixed length buffered chunks. This method greedily packs as many translated Huffman codes as possible into a 64-bit chunk and then stores the number contained separately.

\begin{figure}[ht]
    \centering
    \includegraphics[width=\linewidth]{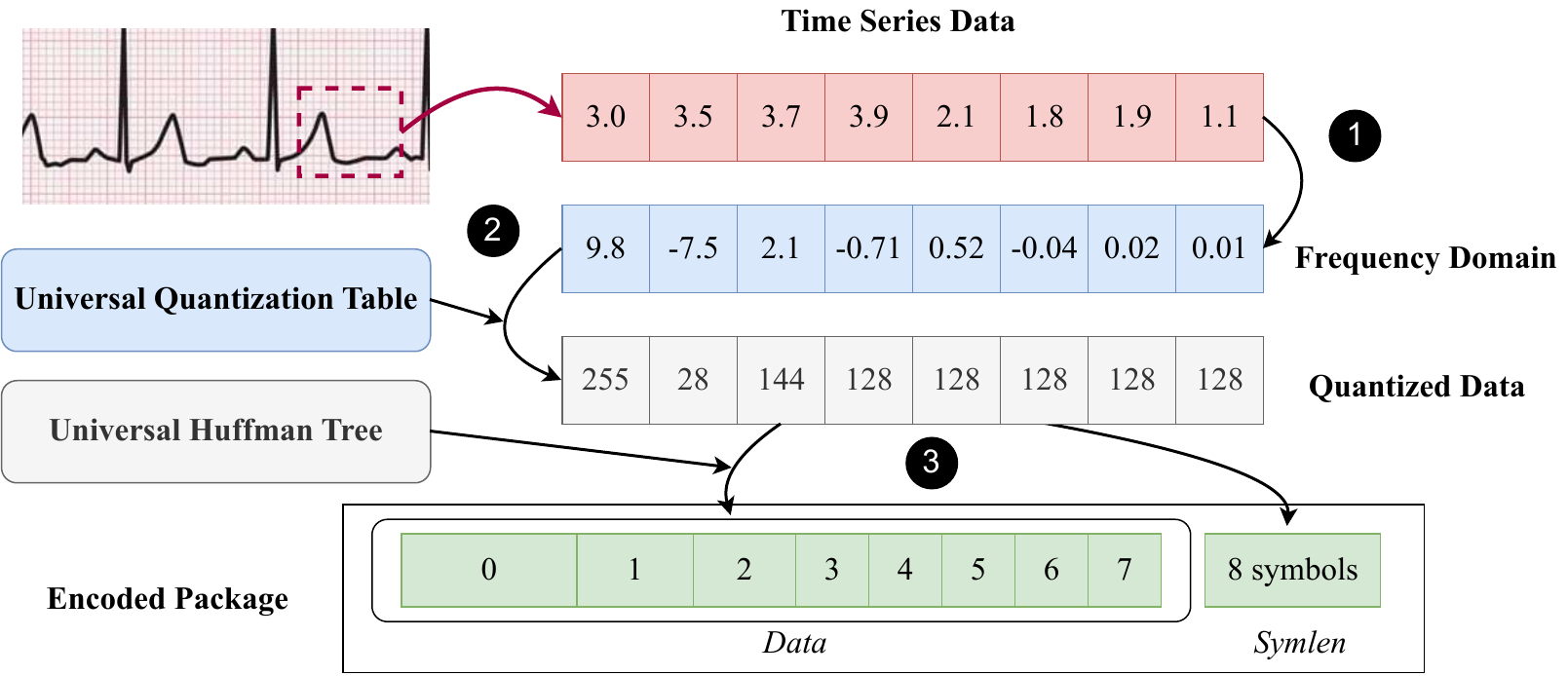}
    \caption{Compression steps on low-complexity encoder including forward DCT-II (1), quantization according to domain table (2) and Huffman \textit{Symlen} word packing (3).}
    \label{fig:f4}
\end{figure}

The algorithm processes the quantized symbol stream sequentially and returns an array of fixed-width 64-bit words. For each symbol, it retrieves the corresponding canonical Huffman codeword and greedily appends that codeword to the current output word if sufficient capacity remains. Otherwise, the current word is finalized, its associated symbol count is recorded in the parallel \emph{SymLen} array, and packing resumes in a fresh 64-bit word. Importantly, codewords are never split across word boundaries: each symbol is either fully contained in the current word or deferred entirely to the next.\\

The \emph{symlen} metadata is the key property of the format. Rather than storing the number of valid bits in each word, \fptc{} stores the number of encoded symbols contained in that word. During decoding, a thread can then stop after reconstructing exactly \emph{symlen[$w$]} symbols, ignoring any padded suffix bits in the tail. This makes each encoded word independently decodable without any inter-thread synchronization, using a prefix-scan of the \emph{symlen} array to index the output of each GPU thread output.

\begin{algorithm}[b]
\caption{Huffman Stream Encoding with SymLen}
\label{alg:encode_symlen_stream}
\begin{algorithmic}[1]
\STATE \textbf{Input:} Huffman map $H$ from symbols to $(code,length)$ pairs; quantized input array $X[0 \dots L-1]$
\STATE \textbf{Output:} encoded 64-bit word array $out[]$ and symbol-count array $symlen[]$

\STATE $i \gets 0$, $w \gets 0$, $count \gets 0$
\STATE $buffer \gets 0$, $bit\_size \gets 0$

\WHILE{$i < L$}
    \STATE $(code, code\_len) \gets H[X[i]]$

    \IF{$bit\_size + code\_len > 64$}
        \STATE $out[w] \gets buffer$
        \STATE $symlen[w] \gets count$
        \STATE $w \gets w + 1$
        \STATE $buffer \gets 0$
        \STATE $bit\_size \gets 0$
        \STATE $count \gets 0$
        \COMMENT{retry the same symbol on next word}
    \ELSE
        \STATE $shift \gets 64 - bit\_size - code\_len$
        \STATE $buffer \gets buffer \;|\; (\mathrm{uint64}(code) \ll shift)$
        \STATE $bit\_size \gets bit\_size + code\_len$
        \STATE $count \gets count + 1$
        \STATE $i \gets i + 1$
        \COMMENT{move to the next symbol}
    \ENDIF
\ENDWHILE

\IF{$count > 0$}
    \STATE $out[w] \gets buffer$
    \STATE $symlen[w] \gets count$
\ENDIF

\STATE \textbf{return} $(out, symlen)$
\end{algorithmic}
\end{algorithm}

\begin{figure*}
    \centering
    \includegraphics[width=\linewidth]{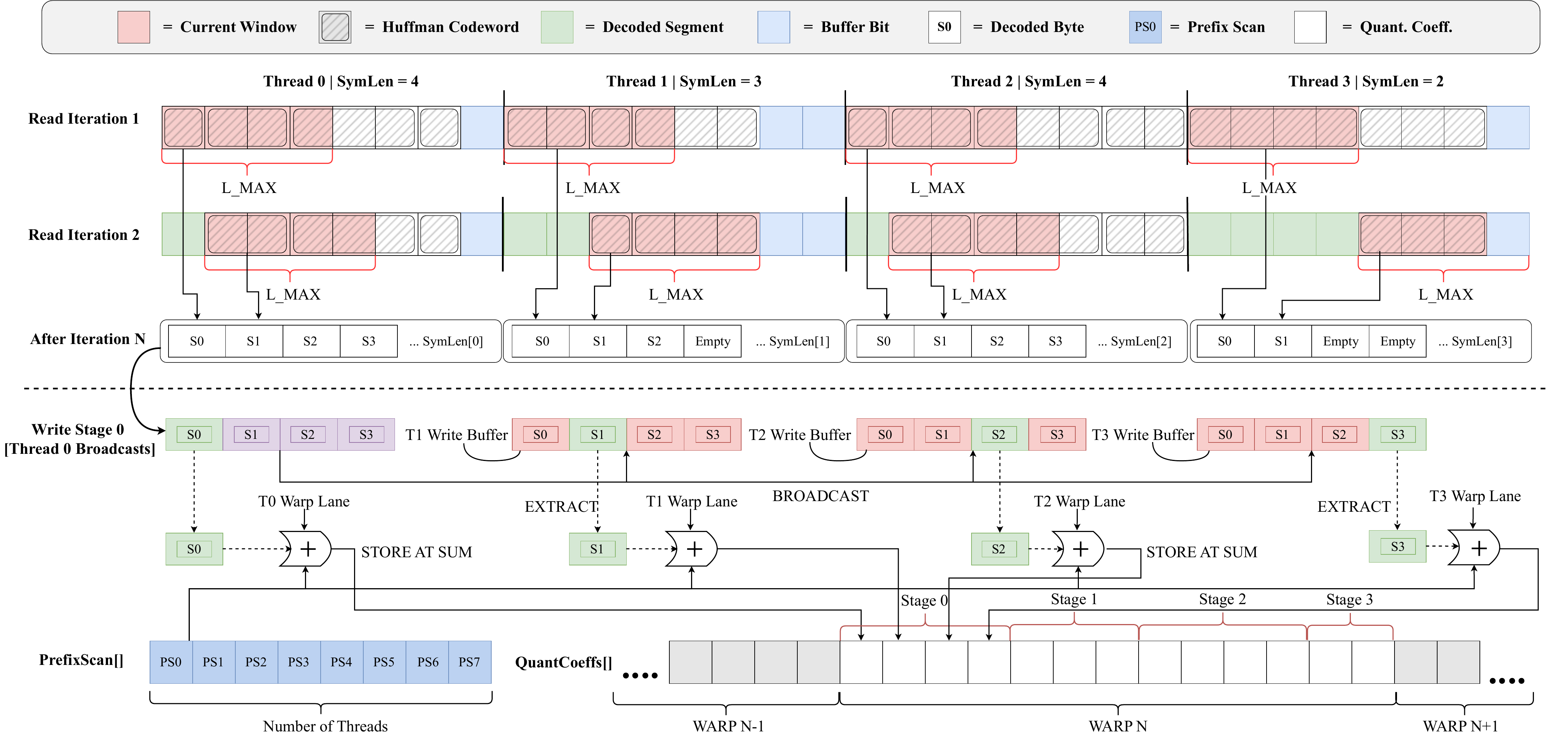}
    \caption{Kernel level diagram of lossless decompression stage with bit-level view of the Huffman code logic using the SymLen method of buffered encodings, and byte-level view of orchestrated thread writes via multistage warp broadcasts.}
    \label{fig:asdf}
\end{figure*}

\subsection{Decompression}

\subsubsection{Lossless} For decompression, \fptc{} must extract the Huffman codewords from the packed bitstream, translate the codeword to a symbol, and reconstruct the time-series format. Paired with the \emph{SymLen} encoder design for 64-bit words, achieving high-throughput memory accesses to the encoded array is straightforward. This is inherent to the work assignment we choose: assigning threads to adopt a fixed-length \emph{encoded} segments, leading to two challenges later on in the decoding. The first challenge being that different threads have different quantities of symbols to decode, and the positioning of all but the first symbol inside of each fixed-length chunks is unknown. Thus, translating their segments in parallel among threads in a warp can lead to repeated divergence. The second challenge is maintaining high-throughput coaslesced memory writes to a unified and fully compacted array due to the variability in the number of symbols each thread decodes.

To address the first challenge, \fptc{} assigns one thread to each 64-bit encoded word and uses the corresponding \emph{symlen} value as the decode termination condition. Each thread therefore decodes exactly the number of symbols stored in its assigned segment rather than searching for delimiters or attempting to infer the end of valid bits. Inside the kernel, the thread repeatedly extracts a prefix of up to $L_{\max}$ bits from its current bit position, indexes a canonical Huffman lookup table, emits the recovered symbol, and advances by the matched code length. Since the code length is bounded, this translation remains constant time per symbol. To reduce lookup latency, the codebook is first staged in shared memory at the block level, and each thread decodes into a small private buffer before performing any global writes. Importantly, threads are not able to determine the bit length of their work assignment prior to decoding, so the buffered bits may be treated as part of a codeword window. Since the codes are prefix-free, a lookup with the buffered bit will still yield the correct value.

The second challenge is compacting the variable-length per-thread outputs into a dense coefficient array. This is resolved by computing an exclusive prefix sum over the \emph{symlen} array before decode, producing the output offset for each compressed word. A naive implementation would then have each thread write its decoded bytes directly to global memory, but such writes are poorly coalesced because neighboring threads generally emit runs of different lengths. Instead, \fptc{} uses a warp-cooperative output stage based on \texttt{\_\_shfl\_sync}: the warp broadcasts one lane's decoded symbols and output offset, and all lanes participate in writing that lane's symbol run to contiguous addresses. This converts irregular per-thread decode output into coalesced global stores while preserving the one-thread-per-word mapping. The result of this first kernel is a fully compacted array of quantized coefficients, which is then consumed directly by the reconstruction kernel.

\subsubsection{Lossy} The output of the lossless decompression stage is an array of spectral quantized coefficients that must have an inverse DCT computed over their float converted values to return to the time-domain representation of the signal data. The choice to separate this into a secondary kernel is motivated by the fact that this stage will have uniform work amongst all threads. This uniform work assignment occurs for dequantization and inverse DCT, so they are fused into one kernel just as Huffman decoding and quantized coefficient compaction are fused, hence the name Dual-Fused Kernel design. A visual representation of the steps on sample data is shown below in Figure 7 at each individual stage of the decompression pipeline.

\begin{figure}[!t]
    \centering
    \includegraphics[width=\linewidth]{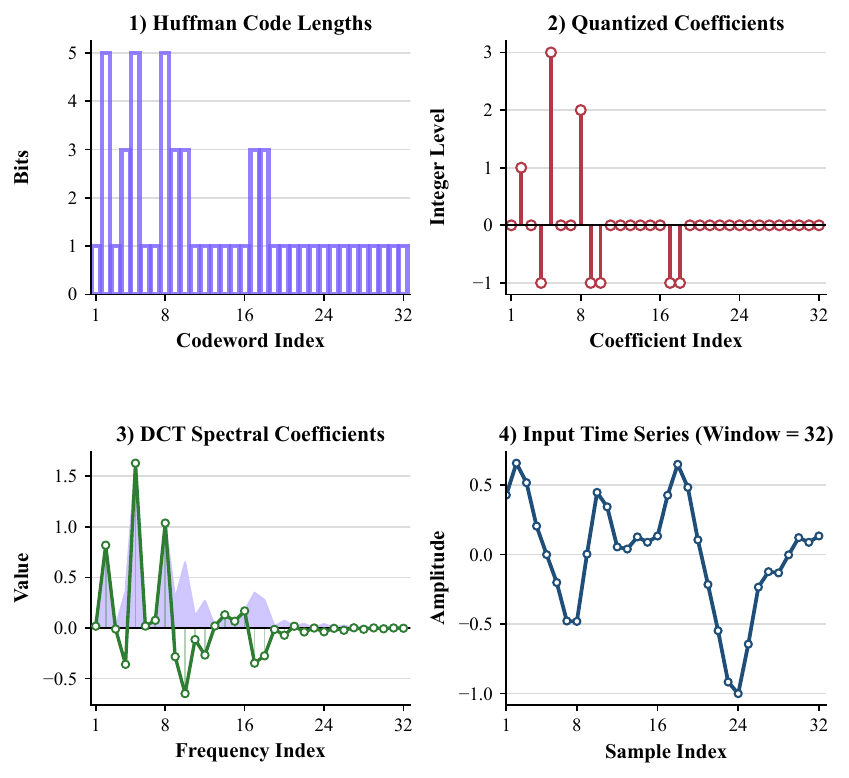}
    \caption{Visualization of data at each decompression step.}
    \label{fig:visual}
\end{figure}

\section{Experimental Setup}

This section outlines the methodology of our experiment including our dataset selection and our three metrics of interest and baseline compression comparisons.

\subsection{Metrics}
We consider the following metrics: compression ratio (CR), percentage root-mean-square difference (PRD), and throughput. Together, these capture the central tradeoffs of lossy signal compression: storage reduction, reconstruction fidelity, and runtime efficiency. Consider $S_{\mathrm{orig}}$ as the original data size and $S_{\mathrm{comp}}$ to be the compressed representation. Let $x_i$ denote the original signal samples and $\hat{x}_i$ the reconstructed samples. Then, the metrics are defined as

\begin{figure}[h]
\centering
\hspace{-1em}%
\begin{minipage}[t]{0.44\linewidth}
\centering\textbf{Compression Ratio}
\end{minipage}%
\hfill
\begin{minipage}[t]{0.52\linewidth}
\hspace{-3em}\centering\textbf{Percentage RMS Difference}
\end{minipage}

\vspace{0.3em}
\hspace{-1em}%
\begin{minipage}[c]{0.44\linewidth}
\centering
\begin{equation}
\label{eq:cr}
CR = \dfrac{S_{\mathrm{orig}}}{S_{\mathrm{comp}}}
\end{equation}
\end{minipage}%
\hfill
\begin{minipage}[c]{0.52\linewidth}
\hspace{-3em}\centering
\begin{equation}
\hspace{-3em}
\label{eq:prd}
PRD = 100 \times\sqrt{\frac{\sum_{i=0}^{N-1} (x_i - \hat{x}_i)^2}{\sum_{i=0}^{N-1} x_i^2}}
\end{equation}
\end{minipage}
\label{fig:metrics}
\end{figure}

Finally, throughput measures the rate at which the decompressor produces output, expressed in GB/s. Since compression operates in real-time, throughput is evaluated only on the decompression side. We measure the time between when compressed data is placed in GPU memory and when the fully decompressed data is written to GPU memory, excluding host-device transfer times, consistent with standard practice for GPU-only data compressors \cite{cuSZp22024}.

\subsection{Datasets}

The evaluation uses ten datasets spanning four signal domains where asymmetric compression is well motivated: biomedical diagnostic, seismic traces, power-grid telemetry, and meteorological sensing. Note that some of these datasets, for example the MIT-BIH Electrocardiogram dataset, are considered canonical yet are rather small studies in terms of data size. To give a fair comparison across compressors that reach maximum throughput at different data sizes, each dataset was duplicated until it exceeded 1 GB in size. Since the duplication is exact, a given set of parameters on any compressor will yield the same CR/PRD performance regardless of if it is run on the original size or the scaled size. 

\begin{table}[H]
\caption{Experimental datasets}
\label{tab:datasets}
\centering
\setlength{\tabcolsep}{4pt}
\begin{tabular}{l l c c}
\toprule
\textbf{Dataset} & \textbf{Domain} & \textbf{Orig (MB)} & \textbf{Scaled (MB)} \\
\midrule
MIT-BIH \cite{mit-bih} & Biomedical & 238.04 & 1160 \\
ECG-ARTH \cite{ecg-raw} & Biomedical & 19.05 & 1000 \\
EEG-MAT \cite{ecg-raw} & Biomedical & 499.88 & 1460 \\
Seismic \cite{seismicdata} & Seismic Traces & 14.98 & 1010 \\
Wind Power \cite{windpower} & Power/Energy & 396.27 & 1160 \\
Solar Power \cite{windpower} & Power/Energy & 396.27 & 1160 \\
Load Power \cite{windpower} & Power/Energy & 396.27 & 1160 \\
Temperature \cite{windpower}& Meteorological & 396.27 & 1160 \\
Irradiance \cite{windpower}& Meteorological & 1160.00 & 1160 \\
Wind Speed \cite{windpower}& Meteorological & 396.27 & 1160 \\
\bottomrule
\end{tabular}
\end{table}

These datasets were selected to cover a wide range of smoothness, sampling characteristics, amplitude distributions, and spectral decay behavior. ECG and EEG represent biomedical waveform domains with strong fidelity requirements. Seismic traces provide a geophysical domain with stricter distortion tolerance and different local structure. Power and meteorological signals represent infrastructure and environmental telemetry, which are often smoother and therefore especially favorable to transform-domain compression.

\subsection{Platform and Baselines}

\fptc{} is evaluated on an NVIDIA RTX PRO 6000 Blackwell Workstation Edition GPU with 96\,GB memory. The evaluation system runs Ubuntu 22.04 with CUDA Toolkit 13. The \fptc{} implementation was compiled with \texttt{nvcc} with -O3, and competing codecs were built according to their public release instructions. The comparison includes recent GPU-oriented lossy compressors that target high throughput and error-bounded reconstruction: cuSZp3 \cite{cuSZp22024}, FZ-GPU \cite{FZGPU2023}, PFPL \cite{PFPL2025}, and cuZFP \cite{ZFP}. Where supported by the implementation, each baseline is tuned to produce points near the PRD values used in the comparison. Because different compressors expose different control parameters and target different error models, the evaluation compares them by achieved PRD rather than by nominal error-bound setting.

\begin{figure*}[ht]
    \centering
    \includegraphics[width=\linewidth]{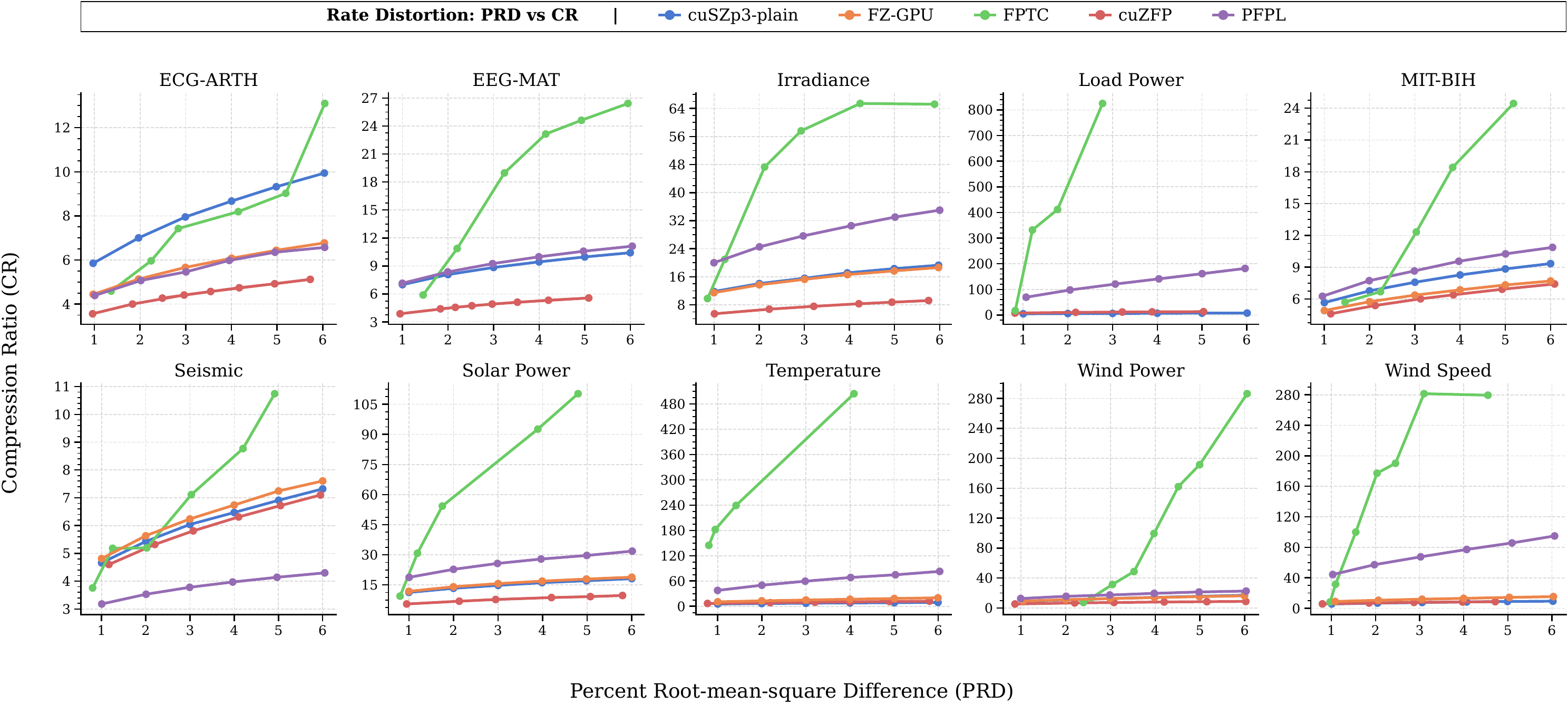}
    \caption{Rate distortion curve performance for all datasets in the PRD range 1--6.}
    \label{fig:rd_curves}
\end{figure*}

\section{Experimental Results}

Our experimental evaluation characterizes the trade-offs between reconstruction fidelity (PRD), compression ratio (CR), and decompression throughput, benchmarking \fptc{} against state-of-the-art GPU-accelerated lossy compressors.

\subsection{Compression}

\subsubsection{Rate Distortion} In signal-processing pipelines, the maximum CR is constrained by the maximum distortion a specific application can tolerate. While this threshold varies across different sensing modalities, empirical studies generally identify up to 5\% as the boundary for high-fidelity reconstruction ~\cite{PRD-ECG, PRD-EEG, PRD-Seismic, Baker2016GMD}. Thus, our rate-distortion curves highlight this range for all evaluated domains.\\

To construct the rate-distortion (RD) curves in Figure~\ref{fig:rd_curves}, each compressor is evaluated over a sweep of its exposed compression controls, and every point is mapped to its $(\mathrm{PRD}, \mathrm{CR})$. For \fptc{}, the sweep is performed over all lossy parameters but focused primarily on the size of the DCT window $N$ and number of retained DCT coefficients $E$. For each signal domain, these parameters are swept over a broad range as an approximation. This produces a dense set of points, which do not exhibit the same tendencies as "single-knob" lossy bounded compressors, hence why a Pareto front is constructed to analyze only the trend of the best suited parameters.

\begin{figure}[h]
    \centering
    \includegraphics[width=\linewidth]{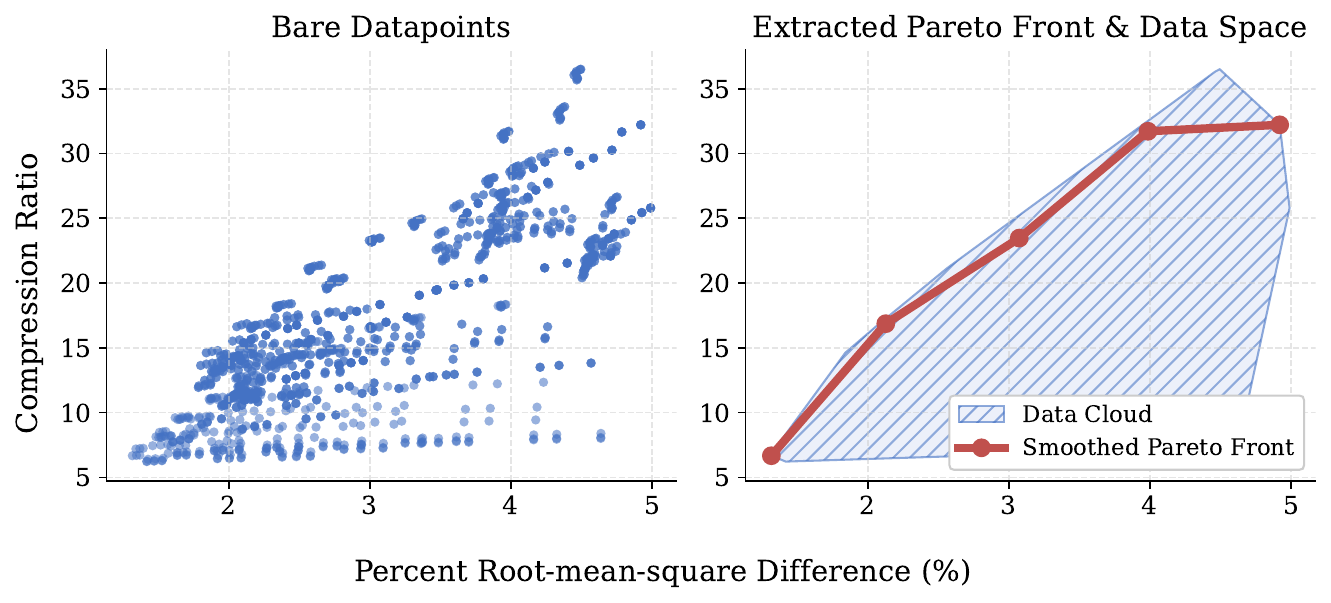}
    \caption{Extraction of Pareto front from a uniform sweep of \fptc{} parameters on a subset of MIT-BIH data.}
    \label{fig:pareto}
\end{figure}

An example of this is shown in Figure \ref{fig:pareto}, where the plot shows compression settings on a sample from MIT-BIH dataset that are uniformly swept, therefore including many impractical configurations that perform poorly and would not be deliberately choosen in a real system.

\subsubsection{Performance Comparison} Across nearly all evaluated datasets, \fptc{} exhibits the strongest Pareto front, achieving substantially higher compression ratios than the baseline compressors at comparable PRD. This advantage is most pronounced on smooth or strongly energy-compacted signals, where \fptc{} achieves dramatically higher compression ratios than all baselines over most of the tested PRD range. In these domains, the DCT truncation is shown to be highly effective, where often the highest compression ratios have only a few retained coefficients.

\observation{In analyzing domain data, \fptc{} parameters can be set not only to target the inherent structure of signal data, but also to target a particular location along the RD curve, mainly by the configuration of parameters $N$ and $E$.}

\fptc{} also performs strongly on biomedical datasets. On MIT-BIH Arrhythmia and EEG-MAT, it consistently exceeds the baselines. The gains are smaller than on the smoothest infrastructure and environmental traces, but still substantial. Notably, the ECG-ARTH dataset is the most competitive case. \fptc{} still achieves the best RD trend, but the margin over the strongest baselines is reduced. Quantifying these improvements via the compression ratio against the strongest competitor per dataset shows clear domain-specific trends: the most significant gains occur in Power/Energy (360\%) and Meteorological (305\%) domains, followed by strong performance in Biomedical (148\%) and Seismic (111\%) data.


\subsubsection{Distortion} The following summarizes acceptable PRD thresholds for domains evaluated in this work based on domain literature.

\begin{itemize}
\item \textbf{Electrocardiography (ECG, target PRD $< 5\%$):} ECG standards establish 5\% as the clinical boundary for excellent reconstruction~\cite{PRD-ECG}, with thresholds up to 10\% considered acceptable before diagnostic reliability degrades. As later discussed, PRD is a global metric and should be interpreted alongside local feature preservation.
\item \textbf{Electroencephalography (EEG, target PRD $< 5\%$):} Clinical validation studies place roughly 5\% as the upper bound for clinician-acceptable reconstruction~\cite{PRD-EEG}, beyond which quality degrades rapidly; PRD above 15\% is generally unacceptable.
\item \textbf{Seismic (target PRD $< 2\%$):} Seismic reconstruction quality is typically reported as NRMSE, with high-fidelity results falling in the 1--2\% range~\cite{PRD-Seismic}. Since PRD is an equivalent normalized RMS measure, this places the high-quality threshold at approximately 2\%.
\item \textbf{Power/Energy (target PRD $\sim < 5\%$):} No standardized PRD threshold exists for this domain. Preserving trends, local ramps, and short-term fluctuations for monitoring and forecasting places the practical operating region in the low single digits~\cite{Tcheou2014SmartGrid}.
\item \textbf{Meteorological (target PRD $\sim < 5\%$):} Similarly lacking a universal threshold, meteorological signals require preservation of seasonal shape and local variations, motivating operation in the low-single-digit PRD range~\cite{Baker2016GMD}.
\end{itemize}

\begin{figure}[h]
      \centering
      \includegraphics[width=1\linewidth]{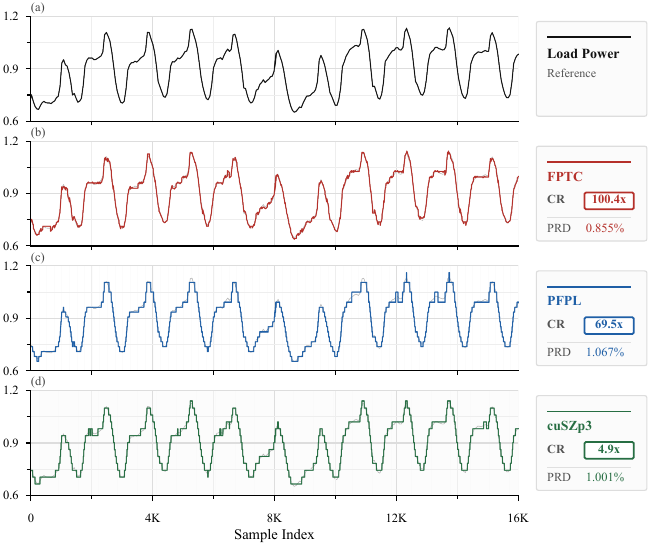}
      \caption{Sample Reconstructed Load Power Data}
      \label{fig:placeholder}
\end{figure}

Since PRD is a global error metric, Figure~\ref{fig:placeholder} provides a qualitative reconstruction comparison on representative load-power data at approximately matched fidelity. Despite sharing PRD, the CR of the three compressors varies significantly from 5x on cuSZp3 to 100x on \fptc{}. It is illustrated that \fptc{} retains a minimal set of coefficients to achieve this results, yet the reconstruction preserves the essential signal structure well.

\observation{Load power data has the largest rate-distortion difference, but even at a given PRD not all compressors preserve local features equivalently, as evident by the block artifacts in the predictive compressors reconstruction. At a certain point, all compressors will suffer from this but the CR limit is different based on the architecture of the compressor.}

At very low PRD (<1\%), some baseline methods match or exceed \fptc{}. This behavior is expected. In the extreme high-fidelity regime, the room for the baseline quantization-aggressiveness used in \fptc{} becomes limited. While full $\mu$-companding can be used to match ultra-low tolerance, it does so at great expense to CR. However, even without this adjustment, \fptc{} remains competitive in the ultra-low distortion regime.

\begin{figure}[h]
    \centering
    \includegraphics[width=.95\columnwidth]{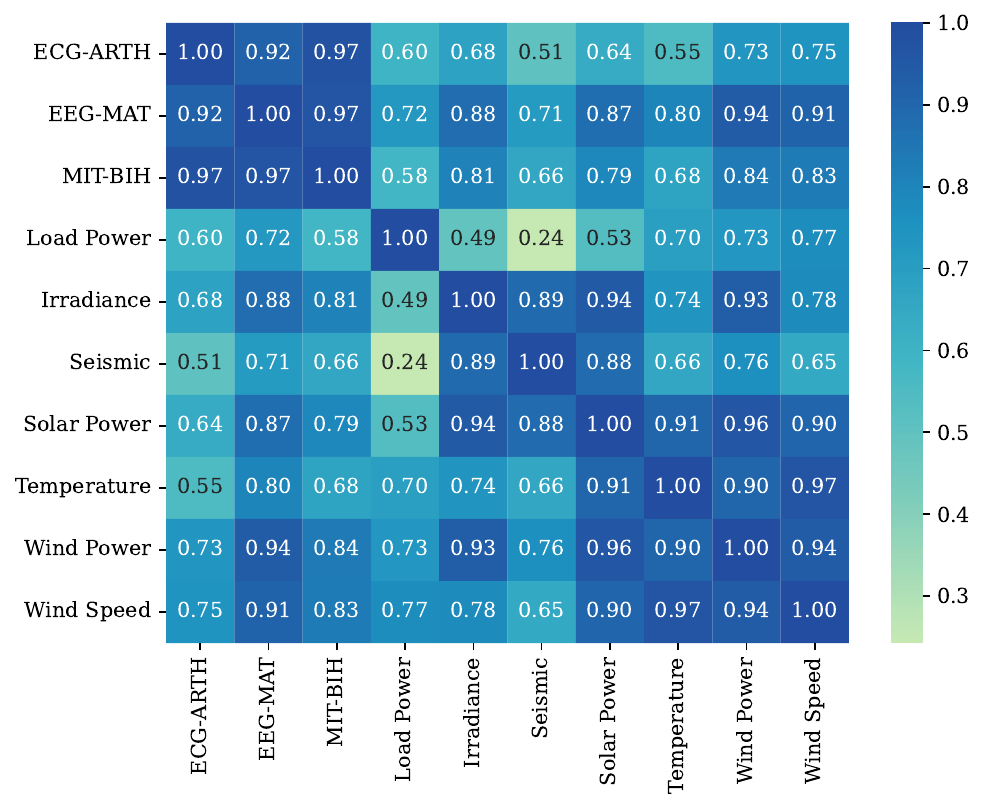} 
    \caption{Correlation matrix between optimized parameters of datasets. Correlation is calculated as the Pearson Correlation Coefficient ($r$). Larger $r$ indicates stronger relationship between the parameters of two datasets}
    \label{fig:fptc_correlation}
\end{figure}

\subsubsection{Dataset Correlation} Another importantant trend observed is the correlation between the optimal parameters for each dataset. As shown in Figure \ref{fig:fptc_correlation}, the optimal compression configurations cluster naturally according to the underlying characteristics of the domain and the sampling frequency of the sensor. For instance, the biosignal domain (ECG and EEG) exhibits an extremely high intra-group correlation ($r \ge 0.92$), suggesting that signals with localized spikes and biological frequency profile converge toward nearly identical Pareto-optimal parameter sets. This result demonstrates that our complex parameters tend to converge for applications within a single domain. This behavior justifies the use of representative domain data to pretrain the compression structures since, for example, structures trained on one ECG study tend to remain performant on across other ECG datasets. Conversely, we find that signals from domains with dissimilar characteristics exhibit distinct optimal parameters. For example, highly non-smooth seismic data requires markedly different parameters than the comparatively smoother Load Power.

\begin{figure*}
    \centering
    \includegraphics[width=1\textwidth]{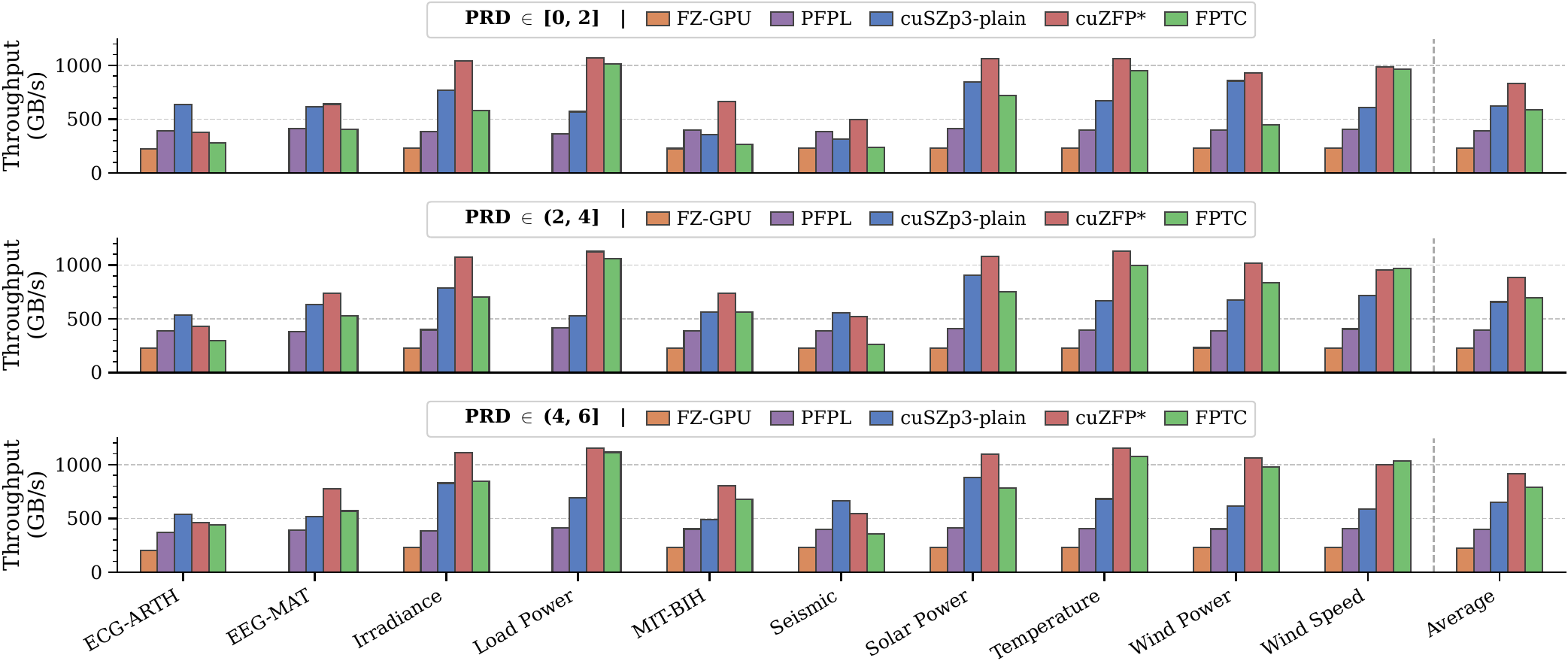}
    \caption{Decompression throughput across Datasets at PRD ranges of $[0,2], (2,4], (4,6]$. Reported throughput taken as average across five trials. *Note that cuZFP is included for comparison despite having unacceptable RD quality on all datasets.}
    \label{fig:throughput_all}
\end{figure*}

\subsection{Throughput}

Unlike the rate-distortion curves, throughput is not plotted over a continuous fidelity axis. Instead, the same parameter sweeps used to generate the RD curves are reused, and every point is grouped by its PRD according to three intervals. This binning enables the comparison of codecs at approximately matched distortion constraints, acknowledging that allowed distortion directly affects throughput. Decode throughputs were measured for a set of parameters in three trials and averaged to account for performance variation of each compressor. Throughput variation is reflected in Table \ref{tab:throughput_by_trial} across five trials on the MIT-BIH Arrhythmia dataset at low PRD tolerance where trials are run immediately in sequence on a GPU that is already warmed.

\begin{table}[h]
    \centering
    \resizebox{\columnwidth}{!}{%
    \begin{tabular}{l ccccc c}
        \toprule
        & \multicolumn{5}{c}{\textbf{Trial \#}} & \\
        \cmidrule(lr){2-6}
        \textbf{Compressor} & \textbf{1} & \textbf{2} & \textbf{3} & \textbf{4} & \textbf{5} & \textbf{Avg.} \\
        \midrule
        FZ-GPU    & 227.20  & 225.50  & 226.29  & 228.21  & 227.83  & \textbf{227.01} \\
        PFPL   & 400.29  & 377.26  & 395.57  & 402.58  & 403.15  & \textbf{395.77} \\
        cuSZp3-plain    & 49.89   & 243.75   & 71.24   & 699.45   & 600.17   & \textbf{332.90}  \\
        cuZFP    & 658.78 & 658.44 & 661.09 & 674.18 & 668.39 & \textbf{664.18}\\
        FPTC     & 243.09 & 288.43 & 243.40 & 243.56 & 295.91 & \textbf{262.88}\\
        \bottomrule
    \end{tabular}
    }
    \caption{Compressor throughput stability (GB/s) on MIT-BIH Arrhythmia dataset at $\text{PRD} \approx 2\%$ across five trials}
    \label{tab:throughput_by_trial}
\end{table}

While cuZFP and FZ-GPU exhibit consist performance, cuSZp3 shows pronounced instability, with throughput varying from $49.89$ GB/s to $699.45$ GB/s across trials. This high variance may pose risks for real-time telemetry pipelines that require predictable latency. \fptc{} remains relatively stable, averaging $262.88$ GB/s, though it exhibits minor jitter in trials 2 and 5. Even with these fluctutations, \fptc{} maintains a reliable performance floor that is sufficient for the requirements of wireless sensor network applications. Across all trials, the throughput of \fptc{} is second only to cuZFP. However, we find that cuZFP exhibits the most significant signal distortion on reconstruction, suggesting that cuZFP emphasizes a high-throughput decoder to the detriment of the achieved compression ratio.

\observation{The throughput of \fptc{} remains competitive across all domains in all PRD ranges despite having a pipeline designed to achieve high rate-distortion performance.}

The measured throughput is especially encouraging on signals with high sampling frequencies. \fptc{} is strongest or near-strongest on Load Power, Temperature, Wind Speed, and Wind Power in multiple PRD ranges. These are the domains in which \fptc{} achieved the largest compression gains, indicating that the same signal structure that improves rate-distortion performance also benefits the GPU reconstruction pipeline in terms of end-to-end decompression throughput. This result is consistent with expectations as memory accesses are dominant the runtime of \fptc{}, and higher compression ratios directly reduce the volume of memory traffic.\\

The most challenging datasets for \fptc{} are ECG-ARTH, EEG, MIT-BIH, and Seismic. Signals from these domains exhibit less smoothness and stationarity, reducing the level of sparsity exploitable by a transform-domain codec. In turn, this makes the reconstruction workload less favorable for \fptc{} because more coefficients are retained, resulting in an increased amount of meaningful work in the lossless stages of the decompressor. Even in these domains, \fptc{} remains within a practical throughput range while offering substantially better compression ratios in comparison to the general-purpose baselines at similar PRD.\\

\clearpage

Another important trend is that \fptc{} generally improves in throughput as the PRD target is relaxed. This is evident across multiple individual datasets as well as in the aggregate averages. The trend is expected: increasingly aggressive compression reduces the amount of retained spectral information, thereby producing less decode and reconstruction work. As a result, \fptc{} often improves compression ratio and decompression throughput together rather than trading one directly against the other.\\

\begin{figure}[ht]
    \centering
    \includegraphics[width=\columnwidth]{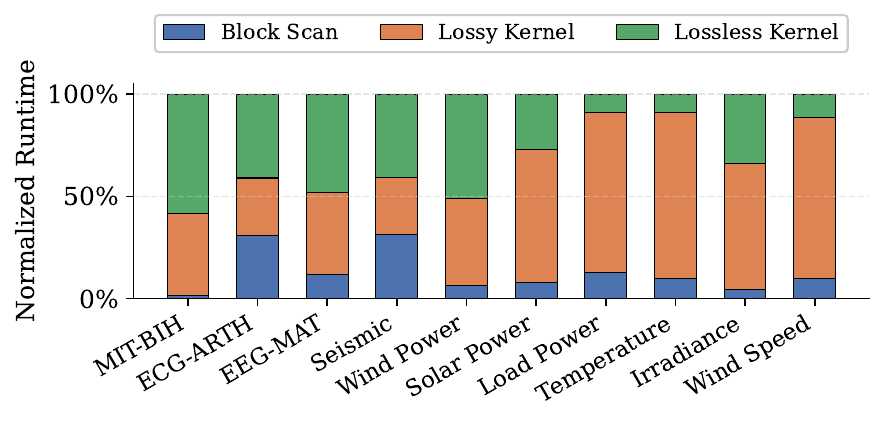} 
    \caption{Normalized runtime breakdown of the three \fptc{} decompression kernels across 10 scaled datasets}
    \label{fig:fptc_breakdown}
\end{figure}

Runtime of individual stages has an interesting pattern. Since the lossless stage processes data from its initial compressed package and expands it up to 8x, the size of data that it handles is considerably smaller in comparison to lossy. However, it is also highly irregular and thus in many cases this adds latency that offsets the reduced read and write sizes. The impact of these varying expansion ratios is quantified in Figure \ref{fig:fptc_breakdown}, which shows kernel latencies across different domains. In datasets like MIT-BIH, the lossless kernel dominates the runtime (60\%), as the signal is not as compresssible as smooth domains. However, on the comparatively smoother Wind Speed dataset, the lossless kernel processes relatively less data, and a high DCT parameter $N$ requires more computations by the lossy kernel to reconstruction the time-domain signal leading to a higher runtime (80\%) in comparison.

This suggests that as signals are increasingly compressed further, the computational cost shifts due to the lossless kernel only having to handled a small number of retained coefficients compared to the large DCT window used to best exploit the frequency sparsity. This indicates that the dual-fused kernel design is optimal for diverse domains here, as a design adapted to one domain would prioritize the work assignment based on the dominant runtime, thus would not generalize across domains at the same throughput.\\

\begin{figure}[H]
    \centering
    \includegraphics[width=\columnwidth]{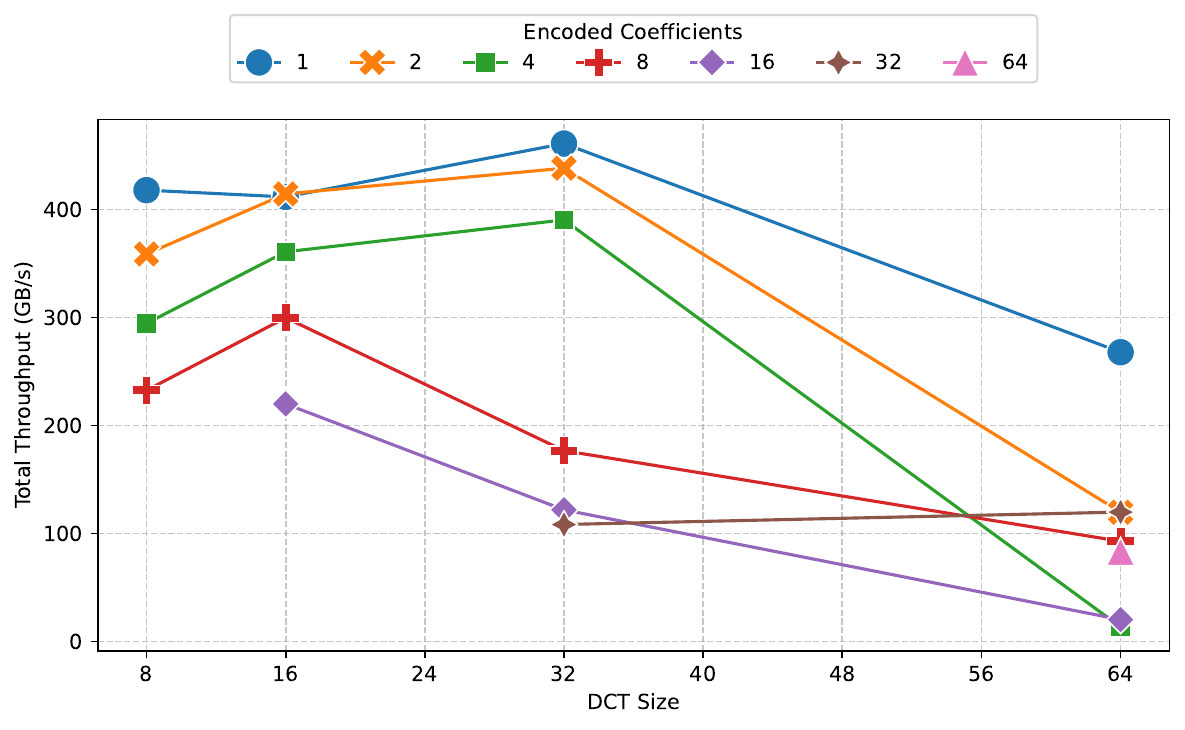} 
    \caption{Throughput as function of DCT, EC on the MIT-BIH dataset. Generally, number of encoded coefficients is inversely proportional to throughput.}
    \label{fig:fptc_dct_ec_throughput}
\end{figure}

Beyond domain-inherent properties, the end-to-end performance is heavily dictated by the specific compression configuration. Specifically, as shown in Figure \ref{fig:fptc_dct_ec_throughput}, throughput is inversely proportional to the number of encoded coefficients (\texttt{ENCODED\_COEFFS}). Fewer coefficients naturally reduce the data volume processed by the Huffman stage and the arithmetic complexity of the lossy kernel. Interestingly, throughput peaks at a \texttt{DCT\_SIZE} of 32 for low \texttt{ENCODED\_COEFFS} values. This operating point reflects a balance between transform overhead and computational complexity, outside of which performance starts to degrade.

It is also important to relate this consideration on throughput back to the rate-distortion curves and visual reconstruction of data. For example, retaining only one coefficient in a 64-point DCT may introduce pronounced block-discontinuity features compared to retaining all coefficients in a 16-point DCT. This will reduce the limit of achievable throughput but we find that high quality reconstruction and embedded-telemtry favorable compression are more important design considerations than maximizing decompression throughput alone.

\section{Summary}

This work presented \fptc{}, an asymmetric lossy codec designed for signal data collected on resource-constrained devices and reconstructed at scale on GPUs. \fptc{} combines a lightweight transform-based encoder with a throughput-oriented GPU decoder built around independent Huffman decoding, compact output placement, and fused reconstruction kernels. The design is motivated by the practical imbalance between low-resource embedded collection and compression and high-volume centralized decompression for archival and analysis purposes.

The evaluation across biomedical, geophysical, infrastructure, and meteorological datasets demonstrates that \fptc{} achieves strong rate-distortion performance while maintaining competitive GPU decompression throughput. In particular, the results demonstrate that transform-domain signal compression, compact entropy-coded representations, and massively parallel reconstruction can be co-designed effectively rather than treated as separate objectives. The correlation analysis further reveals that optimal compression parameters cluster strongly within signal domains, validating the use of domain-representative data to pretrain codec structures that generalize across datasets of the same class. Overall, \fptc{} provides a practical framework for asymmetric compression in real sensing systems where both compression efficiency and large-scale decompression performance are critical.



\clearpage


\bibliography{sample-base}
\end{document}